\documentclass[
	aps,
	prx,
	twocolumn,
	groupeaddress,
	longbibliography,
]{revtex4-2}
\usepackage[english]{babel}
\usepackage[utf8]{inputenc}
\usepackage{amsmath,amssymb,amsfonts}
\usepackage{color}
\usepackage{graphicx}
\usepackage{physics}
\usepackage[normalem]{ulem}
\usepackage{tabularx}

\renewcommand{\selectlanguage}[1]{} % to remove the "language" field from some biblio entries, which gives an error 

\usepackage[
	colorlinks=true,
	linkcolor=blue,
	urlcolor=blue,
	citecolor=blue
]{hyperref}

\begin{document}

\title{
Single-photon emitters in WSe$_2$: Critical role of phonons on excitation schemes and indistinguishability
}

\newcommand{\dtu}{DTU Electro, Department of Electrical and Photonics Engineering, Technical University of Denmark, 2800 Kongens Lyngby, Denmark}

\author{Luca Vannucci}
\email{lucav@dtu.dk}
\affiliation{\dtu}

\author{José Ferreira Neto}
\affiliation{\dtu}

\author{Claudia Piccinini}
\affiliation{\dtu}

\author{Athanasios Paralikis}
\affiliation{\dtu}

\author{Niels Gregersen}
\affiliation{\dtu}

\author{Battulga Munkhbat}
\email{bamunk@dtu.dk}
\affiliation{\dtu}

\date{\today}

\begin{abstract}

Within optical quantum information processing, single-photon sources based on a two-level system in a semiconductor material allow for on-demand generation of single photons. To initiate the spontaneous emission process, it is necessary to efficiently populate the excited state. However, reconciling the requirement for on-demand excitation with both high efficiency and high photon indistinguishability remains a challenge due to the presence of charge noise and phonon-induced decoherence in the solid-state environment. Here, we reconstruct the phonon spectral density experienced by WSe$_{2}$ quantum emitters in the emission process, and we use this information to theoretically analyze the performance of the resonant, phonon-assisted, and Swing-UP of
the quantum EmitteR population (SUPER) swing-up excitation schemes. Under resonant excitation, we obtain an exciton preparation fidelity limited to 0.80 by the strong phonon coupling, which improves to 0.96 for the SUPER scheme (or 0.89, depending on the type of emitter considered). Under near-resonant phonon-assisted excitation, our theory predicts near-unity excitation fidelity up to 0.976 (0.997).
Additionally, we demonstrate that, assuming the suppression of the phonon sidebands, residual dephasing mechanisms such as charge and spin fluctuations are the dominating decoherence mechanisms undermining the photon indistinguishability.

\end{abstract}

\maketitle

\section{Introduction}

Despite remarkable progress over the last four decades, scalable optical quantum information processing continues to be a challenging task on account of the stringent requirements it poses, namely, a source of indistinguishable single-photons is required which simultaneously maximizes the photon efficiency and the degree of indistinguishability towards unity \cite{Heindel2023}.
The former accounts for the number of photons emitted per excitation trigger, whereas the latter is associated with the fact that the interfering photons must be quantum-mechanically identical (with the same polarization, frequency, and spatiotemporal mode).

To date, the most efficient sources of indistinguishable single-photons are based on group III-V semiconductor quantum dots (QDs) \cite{Heindel2023}, with reported efficiency of 0.6 in a lens \cite{Wang2019a} or a fiber \cite{Tomm2021} together with near-unity indistinguishability. 
However, these sources suffer from a challenging integration with different photonic components such as silicon-based waveguides and superconducting detectors, and demand expensive equipment with outstanding control of the growth conditions, comprising techniques such as molecular beam epitaxy and metal organic chemical vapor deposition \cite{Ludwig2017}.

Conversely, layered two-dimensional materials have progressively established themselves as a promising alternative, as they can be readily produced with simple and low-cost mechanical exfoliation \cite{Azzam2021}.
In this context, single-photon emitters in transition metal dichalcogenides (TMDs), such as WSe$_{2}$, have stood out as promising physical platforms due to their unique potential for site-specific strain engineering \cite{Branny2017, Parto2021} and their seamless integration into nanophotonic structures \cite{White2019}. Quantum emitters in WSe$_2$ have recently reached efficiency of 0.65 when coupled to an open cavity \cite{Drawer2023}, thereby rivaling the performance of QD-based emitters in terms of efficiency.
However, indistinguishable single-photon generation from WSe$_{2}$ emitters---and, more generally, from layered materials---has not yet been demonstrated and remains a fundamental challenge \cite{Azzam2021, Fournier2023, Drawer2023}.

A key element in determining both the total source efficiency and the single-photon indistinguishability is the excitation scheme.
Indeed, near-unity exciton preparation fidelity is necessary in order to obtain the maximum total efficiency.
The latter is determined by many concurrent factors, including, among others, the state preparation fidelity, the coupling to a possible cavity mode, and the optical setup used for collection.
Pumping in the \textit{p}-shell guarantees 100\% excitation of the quantum emitter (QE), however it results in poor indistinguishability due to the time-jitter effect \cite{Kiraz2004}. On the other hand, resonant excitation guarantees near-unity state preparation fidelity and indistinguishability \cite{Somaschi2016} at the expense of a reduced total efficiency, due to the cross-polarization setup which is needed to reject the pump.
To circumvent this trade-off, near-resonant schemes such as phonon-assisted exciton preparation have been proposed and implemented \cite{Cosacchi2019, Gustin2020, Thomas2021}. Very recently, exciton preparation close to 100\% has been demonstrated using the two-color Swing-UP of
the quantum EmitteR population (SUPER) scheme \cite{Bracht2021, Karli2022}.
It is well-known that phonon scattering has an important influence on all of these excitation strategies \cite{Reiter2014, Luker2019, Vannucci2023}, and is crucial to understand the emission properties of QD-based emitters \cite{IlesSmith2017}. However, this is largely unexplored for TMD-based emitters, despite several experiments reporting evidence of strong electron-phonon coupling in WSe$_2$ \cite{Li2020, Jin2017, Chow2017}.
Only very recently did the first experimental reports of phonon-related effects on WSe$_2$ QEs appear in the literature \cite{Mitryakhin2023, VonHelversen2023}.
To this day, a quantitative theoretical description of the strength of the exciton-phonon coupling in WSe$_2$ QEs remains elusive. Moreover, the influence of phonons on the source figures of merit, such as the exciton preparation fidelity, the total photon efficiency, single-photon purity, and the indistinguishability requires further investigation.

In this paper, we develop a microscopic description of exciton-phonon coupling in WSe$_2$-based emitters from experimental data.
We achieve this by fitting the prediction of an independent boson model, including the coupling of a two-level emitter to 2D acoustic phonons, to the emission spectrum from strain-localized WSe$_2$ emitters, thereby retrieving the appropriate phonon spectral density.
The latter is then input into a numerically exact tensor network method \cite{Strathearn2018}, which ultimately yields predictions on the excitation dynamics of the system under different schemes---namely, resonant excitation, phonon-assisted pumping, and SUPER swing-up.
We also discuss the fundamental limitations dictated by phonon scattering to the single-photon indistinguishability by analyzing the phonon-induced broadening of the zero-phonon line (ZPL) within a simple pure-dephasing picture.

This paper provides fundamental insights into the physics of phonon coupling in WSe$_2$ emitters, offering an avenue for the precisely tailored control of the excitation process.
Our findings pave the way for developing coherent single-photon sources in TMD materials \cite{Drawer2023}, a crucial advancement in the next generation of photonic quantum technologies.

\section{Phonon coupling from experimental data}
\label{sec:fit}

\begin{figure*}
    \centering
    \includegraphics[width=\linewidth]{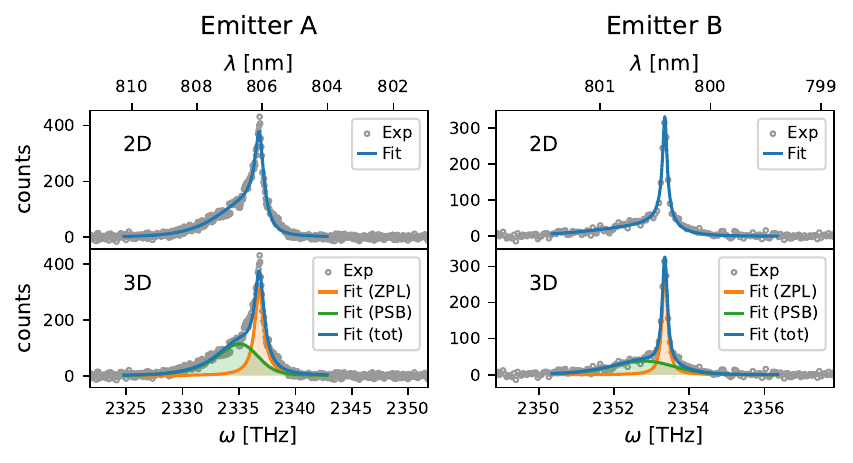}
    \caption{Experimental data from $\mu$PL spectroscopy and fitted theoretical curves from Eq.\ \eqref{eq:spectrum} for a 2D and 3D phonon dispersion at temperature $T = 4$ K. Data are acquired with an integration time of 0.25 s. For the 3D case, the spectrum is separated into ZPL and PSB contributions.}
    \label{fig:fig1}
\end{figure*}

We begin by presenting the photoluminescence spectroscopy of our WSe$_2$ nanowrinkle QEs that typically features a high degree of polarization and high purity \cite{Paralikis2024} and by developing a suitable theoretical description to match their emission spectra.
To obtain localized single-photon emission from WSe$_2$, we exfoliate monolayer and bilayer flakes from bulk and deposit them on top of a dielectric nanopillar as discussed in the literature \cite{Branny2017, PalaciosBerraquero2017, Parto2021, Paralikis2024}.
The combination of out-of-plane strain and native defects localizes the exciton and enables radiative decay from an otherwise dark exciton state \cite{Linhart2019}. To investigate the individual emission from some representative localized WSe$_{2}$ QEs, we performed a micro-photoluminescence ($\mu$PL) measurement under above-band excitation with a 532 nm femtosecond pulsed laser (80 MHz) in a closed-cycle optical cryostat equipped with a microscope objective (60x, NA = 0.82), operating at a temperature of 4 K. Additional details on the fabrication and characterization are provided elsewhere \cite{Paralikis2024, Piccinini2024}.
Experimental data obtained with integration time $\Delta t = 0.25$ s (to minimize the effect of spectral wandering \cite{Paralikis2024}) are reported in Fig.~\ref{fig:fig1} with gray traces.
We find signatures of single-photon emission around 806 nm (emitter A) and 800.5 nm (emitter B). Both peaks show a pronounced spectral asymmetry, which we attribute to phonon sidebands. 

Phonon effects on the spectral properties are investigated within an independent boson model including both a linear \cite{Mahan2000, Nazir2016} and a quadratic \cite{Muljarov2004, Reigue2017, Tighineanu2018} coupling term to longitudinal acoustic (LA) phonons, see Appendix \ref{sec:appA} for details.
The main effect of the linear coupling is to generate a broad phonon sideband (PSB).
Such a term can be conveniently diagonalized using the polaron transformation \cite{Nazir2016, IlesSmith2017}.
On the other hand, the quadratic term induces virtual electronic transitions to higher-lying confined states, thereby broadening the zero-phonon line (ZPL) via a pure dephasing effect at a rate $\gamma_{\rm pd}$ \cite{Reigue2017, Tighineanu2018}.
Furthermore, we use a master equation in the Lindblad form to account for spontaneous decay of the QE at a rate $\Gamma$ and for other sources of pure dephasing, e.g., charge noise at a rate $\gamma_{\rm noise}$. 
We finally fit the experimental data with the function
\begin{equation}
    \label{eq:spectrum}
    \Sigma(\omega) = A\
    \mathrm{Re} \qty[\int_0^{+\infty} \dd{t} e^{\Phi(t)} e^{-i(\omega - \widetilde \omega_X) t} e^{-W t / 2}]
\end{equation}
where $W = \Gamma + \gamma_{\rm pd} + \gamma_{\rm noise}$ and $\widetilde \omega_X$ is the exciton frequency including the polaron shift \cite{Nazir2016}.
The function $\Phi(t)$ is 
\begin{multline}
\label{eq:Phi}
    \Phi(t) = \int_0^{+\infty} \dd{\omega} \frac{J(\omega)}{\omega^2} \qty[ \frac{\cos(\omega t) - 1}{\tanh(\beta \hbar \omega / 2)} - i \sin(\omega t) ]
\end{multline}
with $\beta = (\kappa_{\rm B} T)^{-1}$ the inverse temperature, and depends on the phonon spectral density 
$J(\omega) = \sum_{\vb k} |g_{\vb k}|^2 \delta(\omega - \omega_{\vb k})$, with $g_{\vb k}$ the exciton-phonon coupling constants and $\omega_{\vb k}$ the phonon frequencies.

The most appropriate theory for phonon-coupled QEs in mono- and bilayer WSe$_2$ involves the continuum of 2D acoustic phonons modes in a single layer of WSe$_2$. However, we also consider the case of a 3D phonon dispersion. Assuming spherical electron and hole confinement with wave function $\psi(\vb r) \propto \exp[-|\vb r|^2 / (2R^2)]$ (and identical localization radius $R$ for electrons and holes for simplicity), the phonon spectral density is then
\begin{equation}
\label{eq:phonon_spectral_density}
    J(\omega) = \alpha \omega^z \exp \qty(-\frac{\omega^2}{\omega_{\rm c}^2}) 
\end{equation}
with $z = 2$ or $z = 3$ depending on the dimensionality. The parameter $\omega_{\rm c} = \sqrt{2}c/R$ is a cutoff frequency depending on the electron and hole confinement radius $R$ and the speed of sound $c$, whereas $\alpha$ is given by 
\begin{equation}
    \alpha_{\rm 2D} = \frac{(D_e - D_h)^2}{4 \pi \hbar \rho_A c^4} 
\end{equation}
in 2D with $D_e$ ($D_h$) the electron (hole) deformation potential, and $\rho_A$ the material area density. A detailed calculation is presented in Appendix \ref{sec:appB}. Similarly, we obtain
\begin{equation}
    \alpha_{\rm 3D} = \frac{(D_e - D_h)^2}{4 \pi^2 \hbar \rho_V c^5} 
\end{equation}
in 3D, with $\rho_V$ the volumetric density. Note that $\alpha$ has different physical units in 2D and in 3D.
In the fitting routine, the temperature is fixed at $T = 4$ K, while $\widetilde \omega_X$, $A$, $W$, $\alpha$, and $\omega_{\rm c}$ are free parameters.

\begin{table*}
    \centering
    \begin{tabular}{|c|c|c|c|c|c||c|c|c|}
        \hline 
        & $\widetilde \omega_X$ & $A$ & $W$ & $\alpha$ & $\omega_{\rm c}$ & $R$ & $S$ & $B$ \\
        \hline \hline
        Emitter A (2D phonon dispersion) & 2336.80 THz & 344.09 THz & 0.773 THz & 0.297 ps & 3.209 THz & 1.98 nm & 0.84 & -- \\
        \hline
        Emitter A (3D phonon dispersion) & 2336.76 THz & 347.66 THz & 0.986 THz & 0.232 ps$^2$ & 2.345 THz & 2.71 nm & 0.64 & 0.66 \\
        \hline
        Emitter B (2D phonon dispersion) & 2353.35 THz & 53.37 THz & 0.119 THz & 0.274 ps & 1.959 THz & 3.24 nm & 0.48 & -- \\
        \hline
        Emitter B (3D phonon dispersion) & 2353.35 THz & 52.30 THz & 0.165 THz & 0.634 ps$^2$ & 1.106 THz & 5.74 nm & 0.39 & 0.68 \\
        \hline \hline
        InAs QD \cite{Denning2020} & -- & -- & -- & 0.03 ps$^2$ & 2.2 THz & 2.89 nm & 0.07 & 0.95 \\
        \hline 
    \end{tabular}
    \caption{Fit parameters $\widetilde \omega_X$, $A$, $W$, $\alpha$, and $\omega_{\rm c}$ for emission spectra from emitters A and B. The calculated localization radius $R$, Huang-Rhys factor $S$, and Franck-Condon factor $B$ are also reported, see text for details. Both 2D and 3D phonon dispersions are considered. Typical values for InAs QDs are also reported for comparison from Ref.~\cite{Denning2020}.}
    \label{tab:tab1}
\end{table*}

The 2D and 3D theories are fitted to experimental data for both emitters A and B, with results shown in Fig.~\ref{fig:fig1} and optimal parameters reported in Table \ref{tab:tab1}.
We find good agreement between theory and experiment in all four cases, with a remarkably large value of $\alpha$ as compared with typical InAs QDs values.
Whereas a direct quantitative comparison of the 2D and 3D theory with InAs QDs might be misleading due to dimensional reasons, the dimensionless Huang-Rhys factor $S$ allows for a fair comparison of different scenarios.
Considering only the contribution from LA phonons, the latter is defined as
\begin{equation}
    S = \sum_{\vb k} \frac{|g_{\vb k}|^2}{\omega_{\vb k}^2} = \int_0^{+\infty} \dd{\omega} \frac{J(\omega)}{\omega^2}.
\end{equation}
Using Eq.\ \eqref{eq:phonon_spectral_density} we obtain $S_{\rm 3D} = \frac 1 2 \omega_{\rm c}^2 \alpha$ and $S_{\rm 2D} = \frac{\sqrt{\pi}}{2} \omega_{\rm c} \alpha$. Upon insertion of the best fit parameters, we find $S$ in the range 0.6--0.8 for emitter A and 0.4--0.5 for emitter B, which is one order of magnitude larger than InAs QDs (see Table \ref{tab:tab1}).
This is a quantitative signature of strong exciton-phonon coupling in WSe$_2$, consistent with previous theoretical and experimental work \cite{Wang2019_Raman, Mitryakhin2023}.

While both 2D and 3D fits reproduce the experimental data satisfactorily, the 2D theory appears to be more appropriate for a physical interpretation. Here, it is instructive to compare the value of $\alpha_{\rm 2D}$ obtained from the fit with a prediction based on first-principles calculations. Using $D_e = -6.03$ eV and $D_h = -0.16$ eV for the deformation potentials \cite{Wiktor2016}, $c = 4494 $ m\,s$^{-1}$ for the speed of sound \cite{Haastrup2018, Gjerding2021}, $\rho_A = (m_{\rm W} + 2 m_{\rm Se})/A_{\rm cell}$ and $A_{\rm cell} = \frac{\sqrt{3}}{2} a^2$, with $m_{\rm W}$ and $m_{\rm Se}$ the atomic masses of W and Se and $a = 3.319$ Å the lattice constant \cite{Haastrup2018, Gjerding2021}, we obtain $\alpha_{\rm 2D} = 0.275$ ps, which is very close to the best fit values of $\alpha_{\rm 2D} = 0.297$ ps for emitter A and $\alpha_{\rm 2D} = 0.274$ ps for emitter B.
It is worth noting the large cutoff value of $\omega_{\rm c} = 3.2$ THz obtained for emitter A under a 2D fit theory, which is roughly 50\% larger than for InAs/GaAs QDs.
Such a large value could be explained by a particularly tight localization of electrons and holes in the confining potential.
As detailed in Appendix \ref{sec:appA}, the coupling strength of electron and holes with a plane-wave phonon mode of wave vector $\vb k$ is governed by the form factor $F_{\vb k} = \int \dd[3]{\vb r} e^{i \vb k \vdot \vb r} \qty| \psi(\vb r)|^2$ with $\psi(\vb r)$ the electronic wave function, and $F_{\vb k}$ vanishes with increasing radius $R$ due to the rapid oscillations of the plane-wave mode $e^{i \vb k \vdot \vb r}$.
Specifically, we estimate a confinement radius of $R = \sqrt{2}c/\omega_{\rm c} \approx 2$ nm for emitter A (3.2 nm for emitter B).
Such considerations may be helpful to shed light on the microscopic origin of QEs in WSe$_2$, which is still debated \cite{Linhart2019, Dang2020, Parto2021}.

On the other hand, the advantage of the 3D fit function is that it allows us to separate $\Sigma(\omega) = \Sigma_{\rm ZPL}(\omega) + \Sigma_{\rm PSB}(\omega)$ into the individual contribution of ZPL and PSB respectively, as shown in Fig.~\ref{fig:fig1} (see Appendix \ref{sec:appA} for details). With the 3D fit function it is also possible to obtain an estimate of the Franck-Condon factor as \cite{IlesSmith2017}
\begin{equation}
    B = \exp \qty[-\frac \alpha 2 \int_0^{+\infty} \dd{\omega} \omega e^{-\omega^2/\omega_{\rm c}^2} \coth(\beta \hbar \omega / 2)] ,
\end{equation}
which is not possible for the 2D case \cite{Chassagneux2018}.
Here we obtain $B = 0.66$ (emitter A) and $B= 0.68$ (emitter B) at $T=4$ K, as compared with $B = 0.95$ for InAs QD.
The fraction of light emitted via the ZPL, given by $\qty[\int \dd{\omega} \Sigma(\omega)]^{-1} \int \dd{\omega} \Sigma_{\rm ZPL }(\omega) = B^2$, is then 0.44 for emitter A and 0.47 for emitter B, i.e.\ more than half of the light is emitted via the sideband in both cases. 
These observations further demonstrate a strong exciton-phonon coupling for WSe$_2$ QEs.

Finally, it is worth noticing that we obtain a width $W = \Gamma + \gamma_{\rm pd} + \gamma_{\rm noise}$ in the range 119--986 GHz from the fit. However, the spontaneous emission rate $\Gamma$ is estimated to be 0.1--1 GHz from lifetime measurements, while the phonon-induced dephasing $\gamma_{\rm pd}$ is of the order of $10^{-2}$ GHz for these phonon parameters (see discussion in Sec.\ \ref{sec:indist}). This indicates that a substantial broadening is caused by other sources of noise such as charge noise, which could be potentially reduced by implementing electrical contacts.

\section{Influence of phonon coupling on exciton preparation}

We now analyze the consequences of strong phonon coupling on the QE dynamics, with particular reference to the excited state preparation with laser pulses.
Fast and reliable preparation of the exciton state with fidelity as close as possible to 100\% is crucial to realize efficient sources of indistinguishable photons. 
Whereas above-band excitation has been used in connection with TMD emitters to date  \cite{Mitryakhin2023, VonHelversen2023}, resonant or near-resonant excitation is necessary to meet the indistinguishability requirements for scalable quantum computation. Thus, we focus in the following on resonant excitation, LA phonon-assisted excitation \cite{Cosacchi2019, Gustin2020, Thomas2021}, and the SUPER scheme \cite{Bracht2021, Karli2022, Vannucci2023}.

To model pulsed laser excitation, we couple the QE raising operator $\sigma^\dag$ to a set of laser pulses $\{ \Omega_j(t) \}$, $j \in \{ 1, \ldots, N \}$, with Gaussian shape in time, 
\begin{equation}
\label{eq:Omega}
    \Omega_j(t) = \frac{\Theta_j}{\sqrt{\pi} t_p} e^{-(t / t_p)^2} .
\end{equation}
Here $\Theta_j$ is the pulse area, $\Theta_j = \int_{-\infty}^{+\infty} \dd{t} \Omega_j(t)$, and $t_p$ is the pulse duration.
Each pulse is detuned by a frequency $\delta_j$ with respect to the phonon-shifted exciton frequency $\widetilde \omega_X = \omega_X - D$, with $D = \int_0^{+\infty} \dd{\omega} \omega^{-1} J(\omega)$.
In a frame rotating at the bare exciton frequency $\omega_X$, the Hamiltonian is thus
\begin{align}
    H_{\rm pulse}(t) = \frac{\hbar}{2} \sum_{j=1}^N \Omega_j(t) \qty[ e^{-i (\delta_j - D) t} \sigma^\dag + e^{i (\delta_j - D) t} \sigma]
\end{align}
Both resonant and phonon-assisted excitations make use of a single pulse, while two detuned pulses are necessary for two-color excitation via the SUPER scheme \cite{Bracht2021, Karli2022, Vannucci2023} or the red-and-blue dichromatic protocol \cite{He2019, Koong2021, Vannucci2023}.

To obtain the dynamics, we calculate the reduced density operator $\rho(t)$ for the QE with the formalism of time-evolving matrix product operators (TEMPO) \cite{Strathearn2018}, where the influence of the phonon environment is encoded within a tensor network and treated exactly.
This allows for exact numerical results going beyond the limitations of the polaron transformation or the weak-coupling theory, which break down for fast laser pulses and strong phonon coupling respectively \cite{Vannucci2023}.
Calculations are performed with the \textsc{OQuPy} Python package \cite{OQuPy}, using the 2D phonon spectral density as in Eq.~\eqref{eq:phonon_spectral_density} and with parameters $\alpha$ and $\omega_{\rm c}$ as in Table \ref{tab:tab1}.

The exciton preparation fidelity is calculated as the excited state population right after the pulse, namely $P_X = \Tr \qty[\sigma^\dag \sigma \rho(t)] \eval_{t = 3 t_p}$.
In this Section, we neglect the dephasing $\gamma_{\rm pd}$ induced by quadratic phonon coupling and the noise-induced broadening $\gamma_{\rm noise}$.
Furthermore, we suppress the spontaneous decay term in order to focus purely on the excitation dynamics. This is justified by the fact that exciton preparation occurs on a timescale which is much faster than relaxation (a few ps compared with 1--10 ns).

\subsection{Resonant excitation and Rabi oscillations}

The damping of Rabi rotations in semiconductor QDs due to acoustic phonon coupling is well documented both in theory and experiments \cite{Forstner2003, Ramsay2010a, Ramsay2010b, Luker2019}.
However, at cryogenic temperature this effect is only visible at pulse area of 5--10$\pi$ for QDs. Hence, a resonant $\pi$ pulse is relatively unaffected at $T=4$ K and allows for fast exciton state preparation with near-unity probability. This is especially true for fast pulses ($t_p \sim 1$ ps), where fast optical modulation leads to an effective phonon decoupling from the electronic system \cite{Vagov2007}.

\begin{figure}[ht!]
    \centering
    \includegraphics[width=\linewidth]{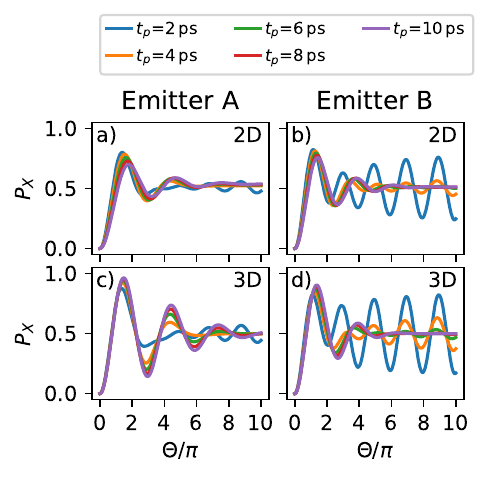}
    \caption{Exciton population $P_X$ after a resonant Gaussian pulse as a function of the pulse area $\Theta$, for an emitter initialized in the ground state. The time duration $t_p$ [see Eq.~\eqref{eq:Omega}] is reported in the legend.}
    \label{fig:fig2}
\end{figure}

Turning to the case of resonant excitation and Rabi oscillations in TMD-based emitters, as depicted in Fig.~\ref{fig:fig2}, we explore the exciton state preparation for emitters A and B with a single resonant optical pulse of variable time length $t_p$. We consider either the 2D or 3D phonon coupling as described in the previous Section, with parameters obtained from the fit.
Both theories predict a noticeable damping of Rabi oscillations due to phonon coupling. However, a striking feature of the 2D prediction is that the first Rabi peak is significantly damped (in the range 0.7--0.8) for both emitters and for all values of $t_p$ considered. 
We conclude that the exciton preparation fidelity is at maximum 80\% under resonant excitation in WSe$_{2}$ due to phonon coupling, in stark contrast with QD-based emitters.
Besides, the total efficiency of WSe$_2$ emitters under resonant excitation is further suppressed by at least a factor of 4 in a cross-polarization setup, due to the fact that they typically possess a single highly-polarized dipole. In contrast, QD-based sources rely on the precession between with two orthogonal V and H dipoles to reduce losses to a factor of 2 \cite{Ollivier2020}.
Finally, it is worth noting that all panels in Fig.~\ref{fig:fig2} show a reappearance of Rabi oscillations at short $t_p$ and larger $\Theta$, a signature of the phonon decoupling effect.

\subsection{Near-resonant phonon-assisted excitation}

To circumvent the need for cross-polarization filtering inherent in resonant excitation, LA phonon-assisted excitation has been used to demonstrate highly efficient sources of indistinguishable single photons based on QDs \cite{Thomas2021}, and to generate three-particle cluster states \cite{Coste2023}.
There, the pump laser is detuned by $\sim$1 THz on the blue side (i.e., around 0.4 nm in wavelength at 900 nm), corresponding roughly to the maximum of the phonon spectral density in Eq.~\eqref{eq:phonon_spectral_density}, and is readily rejected with no additional loss using spectral filtering.
The exciton is then populated via fast phonon-mediated relaxation.
State preparation fidelity reported so far is of the order of 0.90 theoretically \cite{Gustin2020}, and 0.85 experimentally \cite{Thomas2021}.

\begin{figure}
    \centering
    \includegraphics[width=\linewidth]{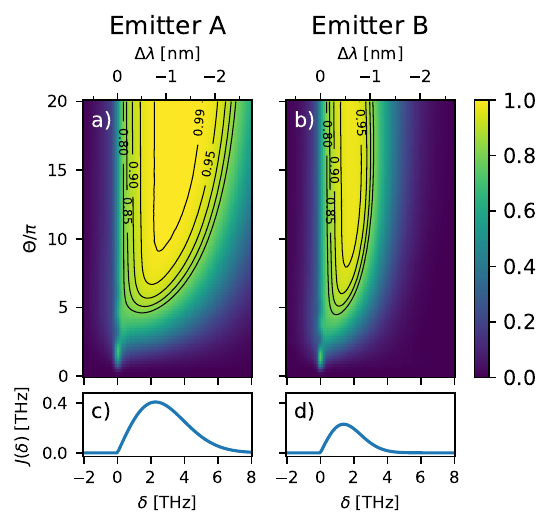}
    \caption{a--b) Exciton population $P_X$ after a detuned Gaussian pulse of duration $t_p = 8$ ps, as a function of frequency detuning $\delta$ and pulse area $\Theta$. The corresponding wavelength detuning $\Delta \lambda$ is also reported, see top axis. c--d) Phonon spectral density $J(\delta) = \alpha \delta \exp \qty(-\delta^2 /\omega_{\rm c}^{2})$ for 2D acoustic phonon coupling as a function of $\delta$. Phonon parameters are as in Table \ref{tab:tab1}.}
    \label{fig:fig3}
\end{figure}

In Fig.\ \ref{fig:fig3} we explore LA phonon-assisted excitation of WSe$_2$ QEs subjected to coupling to 2D acoustic phonons. While sweeping the pulse detuning and amplitude, we observe that near-unity population inversion of $P_X \geq 0.99$ is reached for emitter A over a large parameter range [detuning of 2--4 THz, and pulse area $\Theta \geq 10\pi$, as seen in Fig.\ \ref{fig:fig3}a)], with a maximum of $P_X = 0.997$.
The optimal detuning corresponds roughly to the peak of the phonon spectral density, see Fig.\ \ref{fig:fig3}c).
Emitter B displays a similar qualitative behavior, with $P_X \geq 0.95$ for $\delta$ in the range 1--2 THz and $\Theta \geq 10\pi$, and a maximum of $P_X = 0.976$ [Figs.\ \ref{fig:fig3}b) and \ref{fig:fig3}d)].
Overall, we observe a larger value of $P_X$ as compared with InAs QDs, owing to the stronger nature of phonon coupling in WSe$_2$ emitters.
Moreover, sizable population inversion is obtained at a larger detuning than QD-based sources, making it easier to reject the pumping laser via spectral filtering.
For example, Fig.\ \ref{fig:fig3}a) shows a value $P_X \approx 0.95$ at $\delta \approx 6$ THz (corresponding to 2 nm blue detuning for emission around 800 nm). By comparison, Ref.\ \cite{Thomas2021} reports a maximum $P_X \approx 0.85$ only with rather small detuning of 0.30--0.50 nm, in agreement with theoretical calculations \cite{Cosacchi2019, Gustin2020}.
These observations suggest phonon-assisted excitation as a valuable tool to obtain near-unity excitation fidelity---a necessary condition towards near-unity total efficiency---from WSe$_2$ emitters.

Finally, we note that no population inversion is reported here for $\delta > 8$ THz (emitter A) or $\delta > 4$ THz (emitter B). However, it should be stressed that we are only including the contribution from acoustic phonon coupling in our calculations. A realistic scenario would also include the coupling to optical phonons, which is however expected to show up for $\delta \geq 30$ THz \cite{Mounet2018, Talirz2020, Haastrup2018}, i.e.\ wavelength detuning $\Delta \lambda \leq -10$ nm.
Higher-energy confined states, such as \textit{p}-shell states, are also neglected.
For instance, Kumar et al.~\cite{Kumar2016} identified a blue-shifted exciton (BS-X) at a wavelength detuning of $\Delta \lambda \approx -2.5$ nm (corresponding to $\delta \approx 7.6$ THz), and demonstrated that pumping the ground-state exciton via the BS-X allows for high single-photon purity.
Exciton preparation via the BS-X state, which is not captured by our simple two-level model, will be subject of a follow-up work.

\subsection{SUPER scheme and phonon decoupling}

\begin{figure*}
    \centering
    \includegraphics[width=\linewidth]{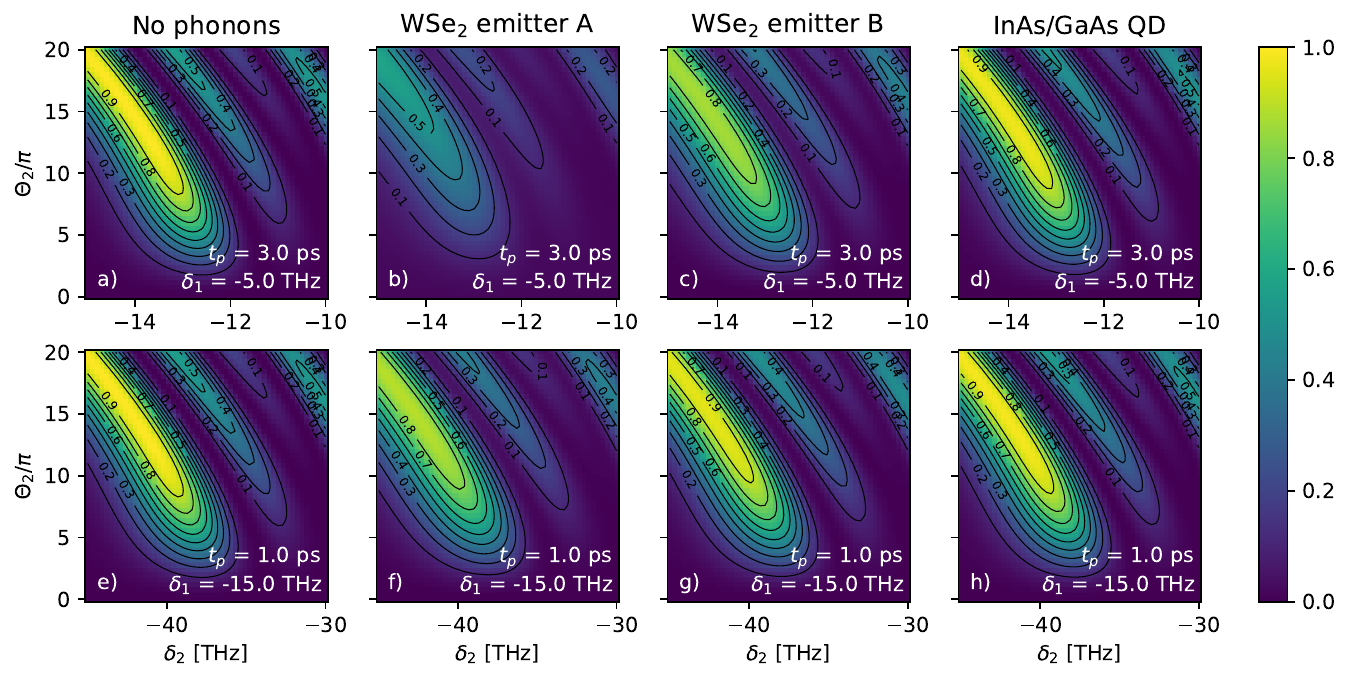}
    \caption{Exciton population $P_X$ after two simultaneous red-detuned Gaussian pulses of identical length $t_p$ but different detuning $\delta_j$ and pulse area $\Theta_j$.
    In panels a--d), the parameters for the closest pulse to resonance are fixed as $\delta_1 = -5.0$\,THz, $\Theta_1 = 11 \pi$, $t_p = 3$\,ps.
    In panels e--h), we fix $\delta_1 = -15.0$\,THz, $\Theta_1 = 11 \pi$, $t_p = 1$\,ps. In all panels, the frequency detuning $\delta_2$ and pulse area $\Theta_2$ of the farthest pulse from resonance are scanned. In a) and e) the phonon coupling is neglected. Elsewhere, we use phonon parameters as in Table \ref{tab:tab1} for emitter A [b) and f)], emitter B [c) and g)], and 3D bulk GaAs [d) and h)].}
    \label{fig:fig4}
\end{figure*}

The SUPER scheme combines the advantages of coherent excitation with a convenient spectral detuning of the laser for straightforward rejection of the pump.
It relies on a coherent swing-up effect of the exciton population generated by two laser pulses acting simultaneously \cite{Bracht2021, Bracht2023}.
An additional practical advantage compared to phonon-assisted excitation is that both pulses are red-detuned with respect to the emitter, which stimulated intense theoretical \cite{Vannucci2023, Heinisch2024} and experimental interest \cite{Karli2022, Boos2024, Joos2023, Torun2023}.
The parameter space generated by two simultaneous pulses is extremely large (including for instance two different pulse detunings $\delta_j$, two pulse amplitudes $\Theta_j$, pulse durations, etc.) meaning that multiple combinations of pulse parameters can activate the swing-up effect.
However, it was shown that the phonon landscape has a strong influence on the latter, generating deviations up to 40\% ($P_X \approx 0.6$) for parameter configurations where $P_X = 1.0$ would be expected in the absence of phonon coupling \cite{Vannucci2023}. 

Here we examine the potential of the SUPER scheme for TMD emitters on the basis of the strong emitter-phonon coupling in WSe$_2$.
In Fig.\ \ref{fig:fig4}, panels a--d), we consider the case of two Gaussian pulses of identical time duration $t_p = 3$ ps, as given by Eq.\ \eqref{eq:Omega}.
The first pulse is detuned by $\delta_1 = -5$ THz, corresponding to 1.7 nm on the red side at 800 nm. The pulse area is set as $\Theta_1 = 11\pi$, and we explore the dependence of $P_X$ on the detuning $\delta_2$ and area $\Theta_2$ of the second pulse.
In this context, it should be noted that ``first'' and ``second'' mean closer and farther from resonance respectively, and have nothing to do with a temporal delay.

A calculation in the absence of phonon coupling [Fig.\ \ref{fig:fig4}a)] shows a bright, extended zone with $P_X = 1.0$ for $\delta_2 \leq -13$ THz, $\Theta_2 \geq 10\pi$.
However, this is reduced to $P_X = 0.59$ and $P_X = 0.87$ at maximum when considering the phonon spectral densities of emitters A [Fig.\ \ref{fig:fig4}b)] and B [Fig.\ \ref{fig:fig4}c)] respectively.
This indicates that the swing-up effect is strongly suppressed due to phonon-induced decoherence, especially for emitter A, which appears to have a stronger coupling to phonons than emitter B (see the discussion of the Huang-Rhys factor in Sec.\ \ref{sec:fit}).
By comparison, we obtain a maximum of $P_X = 0.98$ with identical laser parameters when considering the phonon spectral density of an InAs/GaAs QD coupled to bulk 3D acoustic phonons, as shown in Fig.\ \ref{fig:fig4}d).

In short, a laser configuration which would result in near-unity population inversion for InAs QDs even in the presence of phonons falls well short for the case of WSe$_2$ emitters, due to the stronger phonon coupling affecting the latter.
As discussed in Ref.\ \cite{Vannucci2023}, this problem can be partially solved by adjusting the laser parameters in such a way to effectively decouple the fast swing-up oscillations from the phonon dynamics.
In fact, the dynamics remains qualitatively the same if one applies the scaling $t_p \to t_p / C$ and $\delta_j \to C \delta_j$, with $C$ a constant factor
(this consideration is exact when the dynamics is unitary, i.e., in the absence of phonons).

In Fig.\ \ref{fig:fig4}, panels e--h) we apply this idea with $C=3$, i.e., pushing the value of $|\delta_1| = 15$ THz well beyond the phonon cutoff frequency.
Whereas the dynamics in the absence of phonons remains exactly the same, we observe a general improvement in the population inversion for WSe$_2$ emitters, with a maximum of $P_X = 0.89$ and 0.96 for emitters A and B respectively. The result for InAs QD also improves slightly to 0.99.
Such a value for emitter A is still significantly below unity, indicating that a larger detuning is needed to fully access the decoupling regime.
However, this could be unpractical due to the presence of the optical phonon branch at a detuning of 30 THz (10 nm), which introduces an additional dephasing mechanism via Fröhlich coupling not considered here. Moreover, pumping at such a large detuning may be incompatible with embedding the emitter into narrow-band photonic cavities.

\section{Limitations to single-photon indistinguishability}
\label{sec:indist}

Various factors contribute to reducing the indistinguishability of the emitted photons, such as imperfect or incoherent exciton preparation, charge noise, and phonon scattering.
It seems likely that the established know-how in charge noise control can be transferred from QDs to TMDs with proper engineering of the metallic contacts \cite{Schwarz2016}, and that the broad PSB can be reduced using the cavity filtering effect, as already reported for instance in Ref.\ \cite{Mitryakhin2023}.
On the other hand, broadening of the ZPL induced by quadratic phonon coupling remains an unavoidable limitation \cite{Reigue2017, Tighineanu2018}.
In this section we investigate the fundamental limits to the degree of indistinguishably $\mathcal I$ due to this phonon-induced dephasing mechanism.
Assuming spectral filtering of the sideband, and that the emitter is initially in the $\ket X$ state at $t=0$, the indistinguishability reads
\begin{equation}
    \label{eq:indis}
    \mathcal I = \frac{\Gamma}{\Gamma + 2 \gamma_{\rm tot}} = \frac{1}{1 + 2 \tau \gamma_{\rm tot}}
\end{equation}
with $\Gamma = 1/\tau$ the spontaneous decay rate, and $\tau$ the exciton lifetime. Here, $\gamma_{\rm tot} = \gamma_{\rm pd} + \gamma_{\rm noise}$ includes the contribution $\gamma_{\rm pd}$ induced by the quadratic phonon coupling, and the one from additional noise sources $\gamma_{\rm noise}$.

For the calculation of the phonon contribution $\gamma_{\rm pd}$, we extend in Appendix \ref{sec:appB} the theory developed in Ref.~\cite{Reigue2017} for bulk QDs, obtaining
\begin{equation}
    \label{eq:gamma_pd}
    \gamma_{\rm pd} = \frac{\alpha^2 \mu}{\omega_{\rm c}^4} \int_0^{+\infty} \dd{\omega} \omega^8  \exp \qty(-\frac{2\omega^2}{\omega_{\rm c}^2}) \frac{e^{\beta \hbar \omega}}{(e^{\beta \hbar \omega} - 1)^2}  
\end{equation}
for the case of quadratic coupling to 2D acoustic phonons, where
\begin{equation}
    \label{eq:mu}
    \mu = \pi \hbar^2 \qty(\frac{D_e^2}{\Delta E_e} + \frac{D_h^2}{\Delta E_h})^2 \frac{1}{(D_e - D_h)^4} .
\end{equation}
Here, $\Delta E_i$, with $i \in (e, h)$ is the energy of the first higher-lying state for electrons and holes, which in an isotropic harmonic potential reads $\Delta_i = (\hbar \omega_{\rm c})^2/(2 m_i c^2)$, with $m_i$ the effective masses.
Conversely, $\gamma_{\rm noise}$ is obtained from the fit in Sec.\ \ref{sec:fit} as $\gamma_{\rm noise} = W - \Gamma - \gamma_{\rm pd}$, where $W$ is the total spectral linewidth.
In the following, we focus on the case of emitter A, where phonon-induced limitations are predicted to be larger. 

\begin{figure}
    \centering
    \includegraphics[width=\linewidth]{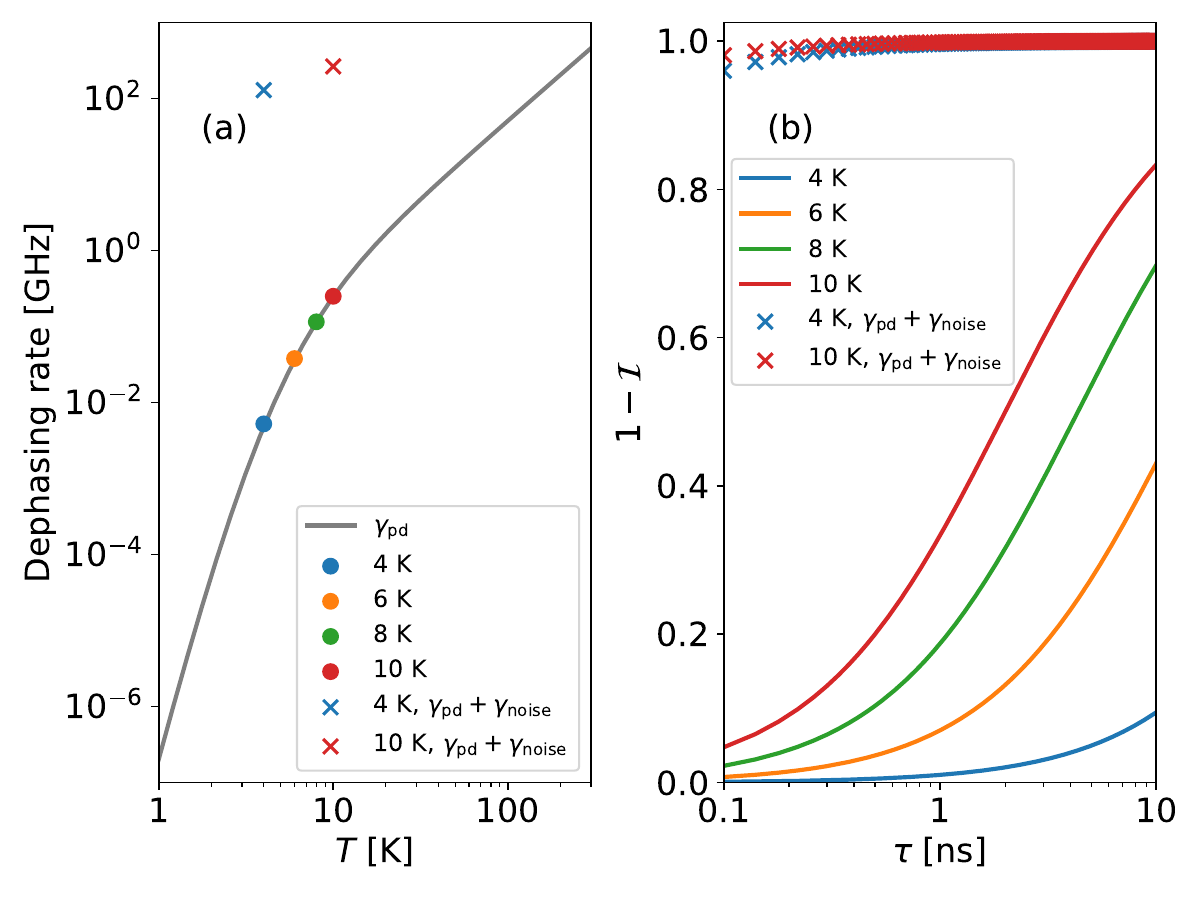}
    \caption{(a) Phonon-induced pure dephasing rate $\gamma_{\rm pd}$ (solid line) for WSe$_2$, calculated from Equation~\eqref{eq:gamma_pd}, as a function of temperature. The filled circles correspond to $\gamma_{\rm pd}$ evaluated at some specific temperatures. The x markers represent dephasing rates extrapolated from fitted $W$ parameters at 4 K and 10 K, where we assume that $\Gamma$ = 1 GHz (i.e., $\tau$ = 1 ns). Deformation potentials are taken from \cite{Wiktor2016}. (b) Indistinguishability in a pure dephasing model as a function of the emitter lifetime $\tau$ (solid lines) for the same set of temperatures from (a), calculated from Equation~\eqref{eq:indis}. As for the x markers, their corresponding indistinguishability was calculated as $\mathcal I = {[1 + 2 \tau (\gamma_{\rm pd} + \gamma_{\rm noise})]^{-1}}$, accounting additionally for the severely detrimental effect of charge noise.}
    \label{fig:figX}
\end{figure}

As depicted in Fig.~\ref{fig:figX}, we obtain a value of $\gamma_{\rm pd}$ of the order of $10^{-2}$ GHz at $T = 4$ K.
The spontaneous emission rate $\Gamma$ is estimated to be 0.1--1 GHz from lifetime measurements. We thus take $\Gamma$ = 1 GHz in Fig \ref{fig:figX} and present a lower bound $\gamma_{\mathrm{noise}} \approx 10^2$ GHz, that is four orders of magnitude larger than $\gamma_{\rm pd}$.
Consequently, the indistinguishability becomes drastically degraded in the presence of both $\gamma_{\rm pd}$ and $\gamma_{\rm noise}$, even at cryogenic temperatures [see Fig.\ \ref{fig:figX}b]. This result is in consonance with the recent findings from Ref.~\cite{VonHelversen2023}, where an estimated two-photon interference visibility of 2\% was reported.

As we show in Fig.\ \ref{fig:figX}b, solid lines, removing $\gamma_{\rm noise}$ allows for $\mathcal I \geq 90\%$ over the entire range of lifetime $\tau \in$ [0.1--10] ns considered. Here, however, indistinguishability is rapidly degraded with increasing temperature, with $\mathcal I$ well below the requirements for scalable quantum information processing already at a temperature of 10 K. Therefore, we suggest that full control of the charge and spin noise at cryogenic temperature is a necessary condition to obtain highly indistinguishable single-photons from WSe$_2$ QEs.

\section{Conclusions}

We have obtained the phonon spectral density for WSe$_2$ QEs coupled to 2D LA phonons by fitting data from $\mu$PL spectroscopy of two representative emitters around 800 nm against the predictions of an independent boson model.
Results show the signature of strong exciton-phonon coupling in WSe$_2$, as quantitatively described by a Huang-Rhys factor which is one order of magnitude larger than for InAs/GaAs QD emitters.

The influence of these findings on different exciton preparation schemes was assessed.
On one hand, strong phonon coupling induces a prominent damping of Rabi rotations, with population inversion limited to 80\% under a resonant $\pi$ pulse. On the other hand, LA phonon-assisted excitation allows near-unity exciton preparation at wavelength detuning of the order of 1--2 nm. 
We have also reported sizable population inversion of 89--96\% under the SUPER scheme, and underlined some fundamental trade-off towards near-unity preparation fidelity dictated by phonon coupling.
We will dedicate a follow-up work to excitation schemes involving higher-energy states, such as the blue-shifted exciton BS-X \cite{Kumar2016}, and optical phonons.

In addition, the limitations on single-photon indistinguishability dictated by phonon coupling were presented under a simple pure-dephasing approximation and assuming filtering of the phonon sideband. Our results point towards additional noise sources, such as charge noise, as being mainly responsible for the lack of indistinguishability at cryogenic temperature reported in the literature so far, and indicate that further investigation of these critical issues in the field is needed.
However, we also show that phonon-induced dephasing deteriorates the indistinguishability via significant broadening of the ZPL already at a temperature of 10 K.

The single-photon indistinguishability was here calculated assuming the emitter to be initially excited with 100\% fidelity. A follow-up work may include a full calculation of the single-photon indistinguishability under different excitation schemes, and within a variational polaron approach \cite{Bundgaard-Nielsen2021} or a tensor network formalism to properly account for phonon coupling.

\begin{acknowledgements}

All authors acknowledge support from the European Research Council (ERC-StG "TuneTMD", grant no. 101076437), and the Villum Foundation (grant no. VIL53033). The authors also acknowledge the European Research Council (ERC-CoG "Unity", grant no.865230),  the European Union’s Horizon 2020 Research and Innovation Programme under the Marie Skłodowska-Curie Grant (Agreement No. 861097).

\end{acknowledgements}

\appendix

\section{Calculation of the emission spectrum with 2D and 3D acoustic phonon coupling}
\label{sec:appA}

Phonon effects on the spectral properties are investigated within an independent boson model making use of the polaron transformation \cite{Nazir2016}. 
The total Hamiltonian is $H = H_0 + H_{\rm ph} + H_{\rm lin} + H_{\rm quad}$ with 
\begin{align}
    & H_0 = \hbar \omega_X \sigma^\dag \sigma , \\
    & H_{\rm ph} = \sum_{\vb k} \hbar \omega_{\vb k} b^\dag_{\vb k} b_{\vb k} , \\
    & H_{\rm lin} = \sum_{\vb k} \hbar g_{\vb k} (b^\dag_{\vb k} + b_{\vb k}) \sigma^\dag \sigma , \\
    & H_{\rm quad} = \sum_{\vb k, \vb k'} \hbar f_{\vb k, \vb k'} (b^\dag_{\vb k} + b_{\vb k}) (b^\dag_{\vb k'} + b_{\vb k'}) \sigma^\dag \sigma.
\end{align}
Here, $H_0$ is the Hamiltonian for a two-level emitter with ground state $\ket G$ and excited state $\ket X$, $\sigma = \dyad{G}{X}$ is the lowering operator, $H_{\rm ph}$ represents the free acoustic phonon bath, whereas $H_{\rm lin}$ and $H_{\rm quad}$ are the linear \cite{Mahan2000, Nazir2016, IlesSmith2017} and quadratic \cite{Muljarov2004, Reigue2017, Tighineanu2018} coupling terms, respectively. The constants $g_{\vb k}$ and $f_{\vb k, \vb k'}$ involve matrix elements of electron-phonon coupling \cite{Reigue2017}.

The linear coupling can be eliminated via the polaron transformation $U_{\rm P} H U^\dag_{\rm P}$, with $U_{\rm P} = \exp [\sigma^\dag \sigma \sum_{\vb k} \omega_{\vb k}^{-1} g_{\vb k} (b^\dag_{\vb k} - b_{\vb k})]$. The transformed Hamiltonian reads
$\widetilde H = \hbar \widetilde \omega_X \sigma^\dag \sigma + H_{\rm ph} + H_{\rm quad}$, where the bare exciton frequency is shifted to $\widetilde \omega_X = \omega_X - D$, with
\begin{equation}
\label{eq:polaron_shift}
    D = \sum_{\vb k} \frac{|g_{\vb k}|^2}{\omega_{\vb k}} = \int_0^{+\infty} \dd{\omega} \frac{J(\omega)}{\omega}
\end{equation}
and $J(\omega) = \sum_{\vb k} |g_{\vb k}|^2 \delta(\omega - \omega_{\vb k})$ is the phonon spectral density.

The microscopic dynamics is described by a second-order master equation for the density operator $\rho$ in the polaron frame, where we include spontaneous decay of the QE with a rate $\Gamma$ and pure dephasing with rate $\gamma_{\rm pd}$. The master equation is \cite{Reigue2017}
\begin{equation}
    \label{eq:ME}
    \dot \rho = -i \comm{\widetilde \omega_X \sigma^\dag \sigma}{\rho} + \Gamma \mathcal L_{\sigma}[\rho] + \gamma_{\rm tot} \mathcal L_{\sigma^\dag \sigma}[\rho]  
\end{equation}
with $\mathcal L_A[\rho] = A \rho A^\dag - \frac 1 2 \acomm{A^\dag A}{\rho}$ the Lindblad dissipator. Here, the total pure dephasing rate $\gamma_{\rm tot} = \gamma_{\rm noise} + \gamma_{\rm pd}$ includes contributions from various sources, such as charge or spin noise and phonon scattering.

The emission power spectrum $\Sigma(\omega)=\langle\sigma^{\dag}(\omega)\sigma(\omega)\rangle$ can be written as
\begin{equation}
    \label{eqn:18}
    \Sigma(\omega)=2\ \mathrm{Re}\Bigg[\int_{0}^{+\infty}\dd{\tau}\int_{0}^{+\infty}\dd{t} e^{-i\omega t} \langle\sigma^{\dag}(t+\tau)\sigma(t)\rangle\Bigg]
\end{equation}
where $\langle\sigma^{\dag}(t+\tau)\sigma(t)\rangle=g^{(1)}(t,\tau)=\Delta(t)\ \tilde{g}^{(1)}(t,\tau)$ corresponds to the first-order dipole correlation function and $\Delta(t)=e^{\Phi(t)}$ is the phonon correlation function.
Using the quantum regression theorem, and assuming that the QE is initially excited at time $t=0$, the correlation function in the polaron frame can be written as $\tilde{g}^{(1)}(t,\tau)=e^{-\Gamma\tau} e^{-\frac{1}{2}(\Gamma+\gamma_{\mathrm{tot}})t} e^{i\widetilde \omega_X t}$. Then, the emission spectrum takes the form
\begin{equation}
    \Sigma(\omega) = \frac{2}{\Gamma}\
    \mathrm{Re}\Bigg[\int_0^{+\infty} \dd{t} e^{\Phi(t)} e^{-i(\omega - \widetilde \omega_X) t} e^{-(\Gamma + \gamma_{\rm tot}) t / 2}\Bigg]
\end{equation}
with $\Phi(t)$ as in Eq.\ \eqref{eq:Phi}.
In Eq.\ \eqref{eq:spectrum} of the main text we substitute the factor $2 \Gamma^{-1}$ with an additional fit parameter $A$ to account for external factors such as the total transmission of the optical setup and the instrument response.

In 3D, thanks to the fact that $\omega^{-2} J(\omega) \coth(\beta \hbar \omega / 2)$ has a finite limit for $\omega \to 0$, the phonon correlation function can be written as $e^{\Phi(t)} = e^{\varphi(t) - \varphi(0)}$ with
\begin{multline}
    \varphi(t) = \int_0^{+\infty} \dd{\omega} \alpha \omega \exp \qty(-\frac{\omega^2}{\omega_{\rm c}^2}) \\
    \times \qty[ \frac{\cos(\omega t)}{\tanh(\beta \hbar \omega / 2)} - i \sin(\omega t) ]
\end{multline}
This observation, together with the fact that $\lim_{t \to \infty} \varphi(t) = 0$, suggests separating the spectrum as $\Sigma(\omega) = \Sigma_{\rm ZPL}(\omega) + \Sigma_{\rm PSB}(\omega)$, with
\begin{equation}
    \Sigma_{\rm ZPL}(\omega) = \frac{2}{\Gamma} B^2\
    \mathrm{Re} \qty[ \int_0^{+\infty} \dd{t} e^{-i(\omega - \widetilde \omega_X) t} e^{-(\Gamma + \gamma_{\rm tot}) t / 2} ] \\
\end{equation}
\begin{multline}
    \Sigma_{\rm PSB}(\omega) = \frac{2}{\Gamma} B^2\
    \mathrm{Re} \left\{ \int_0^{+\infty} \dd{t} \qty[e^{\varphi(t)} - 1] \right. \\
    \left. \times e^{-i(\omega - \widetilde \omega_X) t} e^{-(\Gamma + \gamma_{\rm tot}) t / 2} \right\}
\end{multline}
representing the ZPL and PSB respectively, and where we used $B^2 = -e^{\varphi(0)}$ \cite{IlesSmith2017}.
However, such a separation is not possible in 2D, where $\omega^{-2} J(\omega) \coth(\beta \hbar \omega / 2)$ diverges as $\omega^{-1}$ for $\omega \to 0$.

\section{Calculation of phonon parameters}
\label{sec:appB}

Considering a phonon mode with momentum $\vb k$ and frequency $\omega_{\vb k}$, the matrix element of the electron-phonon coupling are
\begin{equation}
 	M_{\lambda, \vb k}^{ij}
	= \sqrt{\frac{\hbar}{2 m N \omega_{\vb k}}} D_{\lambda} |\vb k| F_{\lambda, \vb k}^{ij}
\end{equation}
with $F_{\lambda, \vb k}^{ij}$ the form factor
\begin{equation}
    F_{\lambda, \vb k}^{ij} = \int \dd[3]{\vb r} e^{i \vb k \vdot \vb r} \psi^{*}_{\lambda, i}(\vb r) \psi_{\lambda, j}(\vb r)
\end{equation}
Here, $D_{\lambda}$ are the deformation potentials, with $\lambda \in (e, h)$ for electrons and holes respectively, $N$ is the total number of ions in the lattice, $m$ the ion mass, and $i$ and $j$ label the electronic states $\psi_{\lambda,i}$ in the confining potential induced by defect and strain engineering.
The exciton-phonon coupling constants $g_{\vb k}$ involve the matrix elements with the ground state wave function $\psi_{\lambda,0}$, and are given by \cite{Nazir2016, Reigue2017, Tighineanu2018}
\begin{equation}
	\hbar g_{\vb k} = M_{e, \vb k}^{00} - M_{h, \vb k}^{00} 
\end{equation}
We assume factorizable wave functions of the form
\begin{equation}
    \psi_{\lambda, 0} (\vb r) = \frac{1}{\sqrt \pi R} \exp \qty(-\frac{x^2+y^2}{2 R^2})  \chi(z)
\end{equation}
with $R$ the in-plane confinement radius and $\int \dd{z} |\chi(z)|^2 = 1$, and take identical confinement length for electrons and holes for simplicity.  
Using a 2D wave vector of the form $\vb k = (k_x, k_y, 0)$ we obtain
\begin{equation}
    F_{\lambda, \vb k}^{00} = \exp \qty(-\frac 1 4 |\vb k|^2 R^2)
\end{equation}
independent of $\chi(z)$.
To calculate the spectral density $J(\omega) = \sum_{\vb k} |g_{\vb k}|^2 \delta(\omega - \omega_{\vb k})$, we turn the sum $\sum_{\vb k} \ldots$ into an integral $\frac{A}{(2\pi)^2} \iint \dd^2{\vb k} \ldots$, assume linear dispersion $\omega_{\vb k} = c |\vb k|$, and introduce $m N = \rho A$, with $A$ the area, and $\rho$ the area density. We finally obtain
\begin{align}
    J(\omega)
    & = \frac{A}{(2\pi)^2} \int_0^{+\infty} \dd{k} \int_0^{2\pi} \dd{\theta} k \frac{ck}{2\hbar \rho A c^2} (D_e - D_h)^2 \nonumber \\
    & \quad \times \exp \qty(-\frac 1 2 k^2 R^2) \delta(\omega - ck) \nonumber \\
    & = \alpha_{\rm 2D} \omega^2 \exp(-\frac{\omega^2}{\omega_{\rm c}^2})
\end{align}
with parameters $\omega_{\rm c} = \sqrt{2} c/R$ and 
\begin{equation}
    \alpha_{\rm 2D} = \frac{(D_e - D_h)^2}{4 \pi \hbar \rho c^4}
\end{equation}

For the calculation of the pure dephasing rate induced by quadratic phonon coupling, we extend the formalism of Ref.\ \cite{Reigue2017} to the case of 2D acoustic phonons. The coupling constants $f_{\vb k, \vb k'}$ are given by
\begin{equation}
\label{eq:f_kk'}
    \hbar f_{\vb k, \vb k'} = \sum_{i \ne 0} \qty( \frac{M_{e, \vb k}^{i0} M_{e, \vb k'}^{0i}}{E_{e,i} - E_{e,0}} + \frac{M_{h, \vb k}^{i0} M_{h, \vb k'}^{0i}}{E_{h,i} - E_{h,0}})
\end{equation}
with $E_{\lambda, i}$ the energy of the $i$-th state, and where the sum involves virtual electronic states with higher electronic states ($i > 0$).
We restrict the sum to include only the first two-fold degenerate excited state $\psi_{\lambda, (1, m)}$, with $m=\pm 1$.
Disregarding the out-of-plane wave function for simplicity, the wave function of the ground state and first excited for a 2D harmonic oscillator are respectively \cite{Cohen-Tannoudji}
\begin{equation}
	\psi_{\lambda, 0}(\vb r) = \frac{1}{\sqrt{\pi} R} \exp \qty(-\frac{|\vb r|^2}{2 R^2})
\end{equation}
\begin{equation}
	\psi_{\lambda, (1, \pm 1)}(\vb r) = \frac{1}{\sqrt{\pi} R^2} \exp \qty(-\frac{|\vb r|^2}{2 R^2}) (x \pm i y)
\end{equation}
A straightforward calculation of the form factor gives
\begin{equation}
    F_{\lambda, \vb k}^{0, (1, \pm 1)} = \frac i 2 R (k_x \pm i k_y) \exp \qty(-\frac 1 4 |\vb k|^2 R^2)
\end{equation}
Upon insertion into Eq.\ \eqref{eq:f_kk'} we obtain
\begin{multline}
    \hbar f_{\vb k, \vb k'}
    = \frac{\hbar}{4 \rho A c} \qty(\frac{D_e^2}{\Delta E_e} + \frac{D_h^2}{\Delta E_h})^2 R^2 \\
    \times \sqrt{|\vb k| |\vb k'|} \qty(\vb k \cdot \vb k') \exp \qty[-\frac{R^2}{4} (|\vb k|^2 + |\vb k'|^2)]
\end{multline}
with $\Delta E_{\lambda} = E_{\lambda, \pm 1} - E_{\lambda, 0}$.
Following Reigue et al.\ \cite{Reigue2017}, the pure dephasing rate induced by quadratic phonon coupling is then
\begin{equation}
    \gamma_{\rm pd} = \Re \sum_{\vb k, \vb k'} \int_0^{+\infty} \dd{s} |f_{\vb k, \vb k'}|^2 \expval{\mathcal B_{\vb k}(s) \mathcal B_{\vb k'}(s) \mathcal B_{\vb k}(0) \mathcal B_{\vb k'}(0) }
\end{equation}
with $\mathcal B_{\vb k}(s) = e^{i c |\vb k| s} b_{\vb k}^\dag + e^{-i c |\vb k| s} b_{\vb k}$
Considering only the contributions from $\vb k \ne \vb k'$, and using $\int_0^{+\infty} \dd{s} e^{\pm i c (|\vb k| - |\vb k'|) s} = \pi c^{-1} \delta(|\vb k| - |\vb k'|)$ we find
\begin{multline}
    \gamma_{\rm pd} = \Re \sum_{\vb k, \vb k'} \frac{\pi}{c} \delta(|\vb k| - |\vb k'|) \\
    \times \qty{n(\vb k) \qty[1 + n(\vb k')] + \qty[1 + n(\vb k)] n(\vb k')}
\end{multline}
Turning the sums into integrals, and using two sets of polar coordinates $(k, \theta)$ and $(k', \theta')$ with both polar axes along the direction of $\vb k$, we finally obtain
\begin{multline}
    \gamma_{\rm pd} = \frac{1}{16 \pi \rho^2 \omega_{\rm c}^4 c^8} \qty(\frac{D_e^2}{\Delta E_e} + \frac{D_h^2}{\Delta E_h})^2 \\
    \times \int_0^{+\infty} \dd{\omega} \omega^8  \exp \qty(-\frac{2\omega^2}{\omega_{\rm c}^2}) \frac{e^{\beta \hbar \omega}}{(e^{\beta \hbar \omega} - 1)^2}  
\end{multline}
which is the same as in Eqs.\ \eqref{eq:gamma_pd} and \eqref{eq:mu} of the main text.

\bibliography{biblio}

\end{document}